\documentclass[12pt,preprint]{aastex}
\begin{document}

\parindent=1.0cm

\title{RAVEN AND THE CENTER OF MAFFEI 1: MULTI-OBJECT ADAPTIVE OPTICS OBSERVATIONS OF 
THE CENTER OF A NEARBY ELLIPTICAL GALAXY AND THE DETECTION OF AN INTERMEDIATE 
AGE POPULATION
\altaffilmark{1,}\altaffilmark{2}}

\author{T. J. Davidge, D. R. Andersen}

\affil{Dominion Astrophysical Observatory,
\\National Research Council of Canada, 5071 West Saanich Road,
\\Victoria, BC Canada V9E 2E7\\tim.davidge@nrc.ca, david.andersen@nrc.ca}

\author{O. Lardi\`{e}re, C. Bradley, C. Blain}

\affil{Department of Mechanical Engineering, University of Victoria, 
\\Victoria, BC Canada V8W 3P2\\lardiere@uvic.ca, cbr@uvic.ca, celia.blain@gmail.com}

\author{S. Oya}

\affil{Subaru Telescope, National Optical Observatory of Japan\\Hilo, HI USA 96720 \\
oya@subaru.naoj.org}

\author{M. Akiyama, \& Y. H. Ono}

\affil{Astronomical Institute, Tohoku University\\6--3 Aramaki, Aoba-ku, Sendai\\
Japan 980-8578\\ akiyama@astr.tohoku.ac.jp, yo-2007@astr.tohoku.ac.jp}

\altaffiltext{1}{Based on data obtained at Subaru Telescope, which is operated by 
the National Optical Observatory of Japan.}

\altaffiltext{2}{This research has made use of the NASA/IPAC Infrared Science Archive,
which is operated by the Jet Propulsion Laboratory, California Institute of Technology,
under contract with the National Aeronautics and Space Administration.}

\begin{abstract}

	Near-infrared (NIR) spectra that have an angular resolution of $\sim 0.15$ 
arcsec are used to examine the stellar content of the central regions of the nearby 
elliptical galaxy Maffei 1. The spectra were recorded at the Subaru Telescope, with 
wavefront distortions corrected by the RAVEN Multi-Object Adaptive Optics 
science demonstrator. The Ballick-Ramsey C$_2$ absorption bandhead near $1.76\mu$m 
is detected, and models in which $\sim 10 - 20\%$ of the light 
near $1.8\mu$m originates from stars of spectral type C5 reproduce 
this feature. Archival NIR and mid-infrared images 
are also used to probe the structural and photometric properties of the galaxy. 
Comparisons with models suggest that an intermediate age population dominates 
the spectral energy distribution between 1 and $5\mu$m near the galaxy center. This 
is consistent not only with the presence of C stars, but also with the 
large H$\beta$ index that has been measured previously for Maffei 1. 
The $J-K$ color is more-or-less constant within 15 arcsec of the galaxy center, 
suggesting that the brightest red stars are well-mixed in this area. 

\end{abstract}

\keywords{galaxies:evolution -- galaxies:elliptical and lenticular, cD -- galaxies: individual (Maffei 1)}

\section{INTRODUCTION}

	Galaxies that host dominant spheroidal components are 
present in large numbers over a wide range of look-back times (e.g. Mortlock et al. 
2013). While spheroids as a group may share some common morphological characteristics, 
this is not indicative of a common evolutionary pedigree. Indeed, in the local 
universe spheroids are likely the result of processes as diverse as mergers 
(e.g. Barnes 1992) and secular evolution (e.g. Kormendy \& Kennicutt 2004).

	Insights into the origins of a spheroid can be gleaned from 
the detailed investigation of its stellar content and structural properties. 
The central regions of galaxies are prime targets for such investigations, as the 
deepest part of the gravitational well harbors signatures of 
past events that played key roles in sculpting the system's present-day appearance. 
Nearby galaxies are of particular interest, as their central regions can be explored 
with intrinsic spatial resolutions that are not possible for more distant objects. 
As a large classical elliptical galaxy in one of the nearest galaxy groups, Maffei 1 
is an important laboratory for examining the evolution of spheroids in moderately 
dense environments. A frustrating complication is that Maffei 1 is observed through 
the plane of the Milky-Way, and efforts to investigate its star-forming history 
(SFH) are confounded by foreground extinction and field star contamination. 

	Buta \& McCall (1999) conducted the first comprehensive survey of the 
photometric and structural properties of Maffei 1 and its neighbors. 
They detected a web of dust lanes that may or may not be 
physically connected to the galaxy. They concluded that (1) 
Maffei 1 has one of the largest angular extents of any galaxy, and (2) 
the light profile follows an $r^{1/4}$ law, making it a `pure' elliptical galaxy.

	Davidge \& van den Bergh (2001) and Davidge (2002) 
observed Maffei 1 with adaptive optics (AO) systems in the near-infrared (NIR) to 
characterize the photometric properties of the brightest resolved asymptotic 
giant branch (AGB) stars. Davidge (2002) found that 
the $K-$band luminosity function (LF) of the brightest stars 
in the outer regions of the galaxy matches that found in NGC 5128, and argued that 
these bright AGB stars are probably among the most metal-rich members of an old 
population. Davidge \& van den Bergh (2005) searched for globular clusters 
belonging to Maffei 1, and measured a cluster specific frequency 
that is comparable to that in other elliptical galaxies 
in low density environments. Candidate blue globular 
clusters were also found, hinting at recent star-forming activity. 
In addition, Davidge (2002) discovered a blue nucleus in the central 
$< 1$ arcsec ($< 16$ pc), which is a source of H$\alpha$ emission 
(Buta \& McCall 2003). Fingerhut et al. (2003) found that 
H$\beta$ in Maffei 1 has an equivalent width of 3.6\AA\ in absorption, which 
is greater than what is seen in the vast majority of nearby 
ealy-type galaxies (e.g. Trager et al. 1998).

	Wu et al. (2014) resolved red giant branch (RGB) and AGB stars in HST 
images of Maffei 1. While the primary goal of their study was to estimate the 
distance to Maffei 1 using the brightness of the RGB-tip, their observations 
also provide insights into stellar content. The brightest AGB stars in their 
(F110W, F110W--F160W) color-magnitude diagram (CMD) of Maffei 1 
have intrinsic brightnesses that are $\sim 1$ magnitude 
fainter in F110W than those in an M31 bulge field (Figure 11 of 
Wu et al. 2014). If the evolved red stellar content of the M31 bulge 
is similar to that of other spheroids (e.g. Davidge 2002), and 
if the brightest stars in the M31 bulge CMD are not disk interlopers, 
then this difference in peak intrinsic stellar brightnesses is 
difficult to explain unless the stellar content of Maffei 1 systematically 
differs from that of other spheroids.

	In the present study, NIR spectra are used to probe the evolved red stellar 
content of the central few arcsec of Maffei 1. These spectra were obtained with the 
RAVEN Multi-Object Adaptive Optics (MOAO: Lardi\`{e}re et al. 2014; Andersen et 
al. 2012) science demonstrator, which feeds turbulence-corrected wavefronts to 
the Infrared Camera and Spectrograph (IRCS: Tokunaga et al. 1998) on the 8 
meter Subaru telescope. The spectra span the $1.4 - 2.3\mu$m 
region and have $\sim 0.15$ arcsec angular resolution. The 
brightest AGB stars contribute significantly to the integrated light 
in the infrared, and the most prominent absorption features in the NIR spectra 
of K and M giants are overtone bands of CO. There are other molecular features 
in this wavelength region, such as the Ballick-Ramsay C$_2$ bandhead near 
$1.76\mu$m, that has the potential to provide powerful age information if detected.

	The spectra are supplemented with archival broad-band NIR and mid-infrared (MIR) 
images that are used to probe the photometric and structural properties of Maffei 1. 
These data bracket the wavelength interval sampled by the spectra, while also probing  
wavelengths where the impact of dust absorption is reduced and the light originates 
from stars that belong to the populations that contribute the bulk of the 
underlying stellar mass. The information obtained from such images provides additional 
clues into the recent evolution of Maffei 1. 

	Various distance estimates for Maffei 1 are summarized in Table 1. The entries 
in this table indicate that Maffei 1 has a distance that is comparable to or -- 
for the most recent estimates -- even lower than that of NGC 5128 (Cen A; $D = 
3.8\pm0.1$ Mpc, Harris et al. 2010). The distance measured by McCall (2014)
hinges on the Humphreys et al. (2013) maser distance determination to NGC 4258, and is 
adopted for calculating spatial scales throughout this paper. For consistency, we 
also adopt the Fingerhut et al. (2003) reddening estimate, which is based on an 
optical depth at $1\mu$m of $\tau_{1\mu m} = 1.69 \pm 0.07$. 

\begin{table*}
\begin{center}
\begin{tabular}{ccc}
\tableline\tableline
Distance & Source & Method \\
(Mpc) & & \\
\tableline
$4.2 \pm 0.5$ & Tonry \& Lupinno 1993 & Surface Brightness Fluctuations \\
$4.4 \pm 0.5$ & Davidge \& van den Berg (2001) & AGB-tip \\
$3.0 \pm 0.3$ & Fingerhut et al. (2003) & Fundamental plane and D$_n - \sigma$ \\
$2.9 \pm 0.3$ & Fingerhut et al. (2007) & Fundamental plane \\
$3.4 \pm 0.3$ & Wu et al. (2014) & RGB-tip \\
$3.3 \pm 0.4$ & McCall (2014) & Fundamental plane \\
\tableline
\end{tabular}
\end{center}
\caption{Distance Estimates}
\end{table*}

	The paper is structured as follows. Details of the observations and the 
processing of the spectra and the images used in the structural 
analysis are presented in Section 2. The isophotal 
properties of Maffei 1 in the NIR and MIR are examined in Section 3, while the spectra 
are investigated in Section 4. A summary and discussion of the results follows 
in Section 5.

\section{OBSERVATIONS}

\subsection{Spectra}

	Spectra of Maffei 1 were recorded with the Subaru IRCS during the 
commissioning of the RAVEN MOAO science demonstrator on UT August 11, 2014. 
Six 300 sec exposures were recorded of Maffei 1. These were the first observations of 
an extragalactic target to be recorded with RAVEN. 
The star HD18077 was observed immediately following Maffei 1 to monitor telluric 
absorption features and also to provide calibration information to remove 
the thermal background. This star is $\sim 2.5$ degrees East of Maffei 1, 
and so was observed at an airmass that is comparable to that of Maffei 1.
With a spectral-type A2V, HD18077 is a suitable template for 
removing telluric signatures without affecting the absorption 
features that are seen in Maffei 1. 

	The corrected field of view (FOV) 
in classical AO systems is limited by isoplanatism -- the light paths of 
objects separated by angular offsets in excess of a few arcsec pass through 
different atmospheric turbulence cells, and so the optimal
correction towards one object differs from that 
towards another. MOAO overcomes this limit on the FOV by separating the
sensing and correction of wavefront distortions. At least three guide
stars are used in an MOAO system to construct a three
dimensional tomographic map of the turbulence across the science field. RAVEN
has three natural guide star wavefront sensors (WFSs) that
patrol a 3.5 arcminute diameter field as well as an on-axis WFS designed 
for use with the Subaru laser guide star; the latter was not 
used during these observations. The delivered level of correction
depends on how well the turbulence is mapped, and future facility MOAO
systems will likely deploy more WFSs over even larger fields
to minimize residual tomographic wavefront error.

	Wavefront correction in MOAO systems is done individually for each
science target. The use of separate optical paths with autonomous 
deformable mirrors (DMs) means that the optimal AO correction can be applied
for each target, and future facility MOAO instruments may have many parallel 
science channels. As an MOAO science demonstrator, RAVEN has just two science 
pick-offs that patrol the 3.5 arcminute field. Wavefront distortions are
corrected by an $11 \times 11$ element DM in each science path. 

	By separating the sensing and correction of atmospheric turbulence, 
MOAO systems must operate in open loop mode. Open loop
control requires the WFSs to have a very large dynamic range; that is, they
must measure the uncorrected turbulence distortions very
accurately, and without the wavefront correction 
feedback that is present in closed-loop systems. Any non-linearities or hysteresis 
in the DM will thus degrade performance, since these sources of error are not sensed. 
While the use of open loop control poses technical challenges,
open loop MOAO instruments benefit from compact system design and
potentially higher throughput than traditional AO and Multi-Conjugate AO systems.

	As a science demonstrator, RAVEN was designed to balance the degree of 
correction with sky coverage on a limited financial budget; the tomographic error, 
determined by the number of WFS pick-off arms, is balanced with the fitting error 
that is set by the number of DM actuators. High Strehl ratios are not 
expected with Raven. Still, significant gains in the encircled energy 
are realised. Future facility MOAO systems could produce diffraction-limited 
performance by employing higher order DMs and a greater density of WFSs. 

	When used in spectroscopic mode, the light from the two pick-offs 
is re-imaged side-by-side by Raven onto the long slit of the IRCS. 
Each science probe feeds a 7 arcsec long segment to the IRCS, with 0.060 arcsec 
pixel$^{-1}$ sampling on the $1024 \times 1024$ ALLADIN III InSb array that is 
the spectroscopic detector. This detector has a dark current $\sim 0.05$ e$^{-}$ 
sec$^{-1}$, and this is roughly two orders of magnitude lower than the signal from 
Maffei 1 detected at the slit ends in $H$.

	For these observations one science pick-off sampled the center of 
Maffei 1, and the other sampled a field along the minor axis 1 arcmin from 
the galaxy center. The light was dispersed with the HK grism, and the spectral 
resolution is $\lambda / \Delta \lambda \sim 700$ with $\sim 11 \AA\ $pixel$^{-1}$ 
sampling. The $K-$band spectrum has a disappointing S/N ratio and does not probe 
the same radial extent as the $H-$band spectra, probably due in part to high levels 
of emissivity from warm optics in RAVEN. The $K-$band spectrum is 
not considered further.

	The individual spectra were combined by taking the median signal at each 
pixel location. The median was selected for image combination as it is a more robust 
statistic in the presence of outliers (e.g. as could result from cosmic-rays) when 
only a modest number of exposures are available. Instrumental and atmospheric 
signatures were then removed to produce a two-dimensional spectrum from which binned 
spectra covering various radial intervals could be extracted (Section 4). This 
involved the following steps. First, the two-dimensional spectrum was divided by a 
flat-field frame constructed from images of dispersed light from a continuum source. 
Next, a thermal background calibration frame was constructed from 
dithered observations of HD18077. This was subtracted from the flat-fielded spectra 
to remove thermal emission from the sky and from warm opto-mechanical elements 
along the optical path.

	Telluric emission lines were then subtracted from the spectrum. Initially, it 
was thought that this could be done by subtracting the light from the science 
pick-off that samples off-nucleus light. However, the throughput 
of this science path was found to differ from that of the pick-off 
that sampled Maffei 1, and residual sky lines remained after differencing. Residual 
sky features were then removed by finding a scaling factor that 
could be applied to the off-nucleus channel that suppressed 
residuals near the edge of the Maffei 1 science pick-off slit segment. The 
resulting sky-subtracted spectrum was then wavelength calibrated using a dispersion 
solution that was obtained from Ar arc lamp observations. The final 
processing step was to remove telluric absorption features. This was done by 
dividing the Maffei 1 spectrum by a spectrum of HD18077 that 
was normalized to unity and that had Brackett absorption lines removed 
after modelling these features with Voigt line profiles.

\subsection{Archival Imaging Data}

\subsubsection{SPITZER Observations}

	Observations of Maffei 1 in [3.6] and [4.5] were recorded 
as part of Spitzer program 57359 (PI: K. Mansi). The Spitzer images were recorded 
over three epochs with a different satellite orientation for each epoch. 
The exposure time is 93.6 sec per epoch, and 
the angular resolution is $\sim 1.8$ arcsec ($\sim 33$ pc) FWHM. 

	Stacked images for each epoch were available 
as Post-Basic Calibrated Data (PBCD) from the Infrared 
Processing and Analysis Center (IPAC) Spitzer Heritage Archive 
\footnote[3]{http://sha.ipac.caltech.edu/applications/Spitzer/SHA/}.
While the central few arcsec of Maffei 1 are saturated in these images, they are 
still of interest because they sample the circumnuclear region at wavelengths 
where much of the light comes from the luminous red stars that also dominate the 
NIR spectrum of Maffei 1. The PBCD images from each epoch were rotated and aligned 
to a common reference frame, and then averaged together. The results were
trimmed to the $\sim 4.9 \times 5.4$ arcmin$^2$ area of common overlap. 
Background light levels were measured $\sim 3$ arcmin from the galaxy 
center along the minor axis of Maffei 1, near the edge of the trimmed images, 
and these background levels were subtracted from the combined images. The 
[3.6] image was convolved with a Gaussian to match the angular resolution 
of the [4.5] image. 

\subsubsection{2MASS Observations}

	$J$, $H$, and $K$ images of Maffei 1 were downloaded from the 
IPAC Large Galaxy Atlas (LGA: Jarrett et al. 2003) host 
\footnote[4]{http://irsa.ipac.caltech.edu/applications/2MASS/LGA/}. 
The LGA images cover a $\sim 23 \times 23$ arcmin$^2$ field, with an angular 
resolution of $\sim 3$ arcsec ($\sim 55$ pc) FWHM. The $H$ and $K$ 
images were smoothed to match the angular resolution in $J$. Background sky levels 
were measured along the minor axis near the field edges, and the results were 
subtracted from the LGA images. 

\section{RESULTS: Broad-Band imaging}

\subsection{Isophotal Properties}

	The structural properties of a galaxy are an important part of its fossil 
record, providing insights into past events that are complementary 
to those gleaned from the spectral-energy distribution (SED). A 
complication for Maffei 1 is that it is obscured by a complex web of dust absorption 
(Buta \& McCall 1999; 2003). Structure in the dust distribution may skew isophotal 
measurements, with the greatest potential impact at wavelengths $< 1\mu$m. This 
being said, the presence of dust and its physical relationship with Maffei 1 is of 
interest since -- if it is associated with Maffei 1, and not the foreground -- 
then it is a possible signature of a recent merger or accretion event. 
In this study, the isophotal properties of Maffei 1 are investigated using the 
archival Spitzer and 2MASS images described in Section 2. The effects of dust 
absorption are greatly reduced at these wavelengths when compared with the visible 
part of the spectrum. In addition, the stars that contribute most of the light at 
these wavelengths originate from the populations that likely dominate 
the stellar mass of Maffei 1.

	Isophotal measurements were made with the
{\it ellipse} program (Jedrzejewski 1987), as implemented in STSDAS.
Foreground stars complicate the fitting of isophotes to Maffei 1. It is 
difficult to remove contaminating stars that are close to the center of Maffei 1  
from the images as they are projected against a bright, non-uniform background. Many 
are saturated, and this renders their removal problematic. Given these difficulties, 
the sigma-clipping option in {\it ellipse} was used to mask regions in 
which residuals deviated from the fit at the $2.5\sigma$ and higher level, as 
experiments indicated that this level of clipping effectively masks stars. 

	If the point spread function (PSF) of a star is not completely removed 
by sigma-clipping then residual signal in the outer wings of the PSF might 
skew the isophotal measurements, although this is mitigated by 
azimuthal averaging. The PSFs of the brightest stars 
can be traced out to a radius of $\sim 9$ arcsec before blending into the background 
noise. Therefore, if the surface brightness measurements are skewed by residual 
signal in the PSF wings then structure on an 18 arcsec angular scale (i.e. twice 
the PSF radius) would be present. 

	The influence that unsuppressed signal from the PSF wings might have on the 
surface brightness measurements was assessed quantitatively using the residual 
images that are produced by subtracting a galaxy model based on 
the fitted isophotes from the initial images. The statistical 
properties of pixel intensities in selected rectangular regions of the 
residual image were examined. The regions for investigation were selected 
(1) based on the absence of stars as determined by visual inspection, and (2) 
because they sample radial distances that exceed those that will be affected 
by residual PSF wing signatures. The dispersions in pixel intensities in 
these regions suggest that faint PSF wings may account for no more than $\pm 0.1$ 
magnitude uncertainty in the surface brightness at the largest radius that was 
photometered in [3.6], and $\pm 0.2$ magnitude in [4.5]. This is an upper limit as 
the dispersion includes photon noise from the sky and the 
main body of the galaxy. An examination of the 2MASS residual images yields 
similar results. We conclude that the isophote-subtracted SPITZER and 
2MASS images are largely free of the distortions that 
would be expected if isophote fitting was skewed by individual stars.

	The surface brightness, ellipticity, and the fourth 
order coefficient of the Fourier expansion of the isophotes -- B4 -- 
found by running {\it ellipse} on the 2MASS $J$ and SPITZER [4.5] images 
are shown in Figure 1. The measurements obtained from the $H$ and $K$ images 
are similar to those in $J$, and so are not plotted. Likewise, the isophotal 
measurements made in [3.6] are very similar to those in [4.5], and so are also 
not shown. The $J$ surface brightnesses were computed using photometric calibration 
information in the image headers while the [4.5] surface brightness measurements 
were calibrated using zeropoints from Reach et al. (2005). 

	The inner few arcsec of Maffei 1 are saturated in [4.5], and this fixes 
the inner radius of measurements in this filter to $r \sim 6$ arcsec. The 
angular resolution of the 2MASS images is $\sim 1.5\times$ poorer than 
that of the SPITZER images. However, the nucleus is not saturated. There is thus 
overlap with the RAVEN observations, albeit with poorer angular resolution.

\begin{figure}
\figurenum{1}
\epsscale{1.0}
\plotone{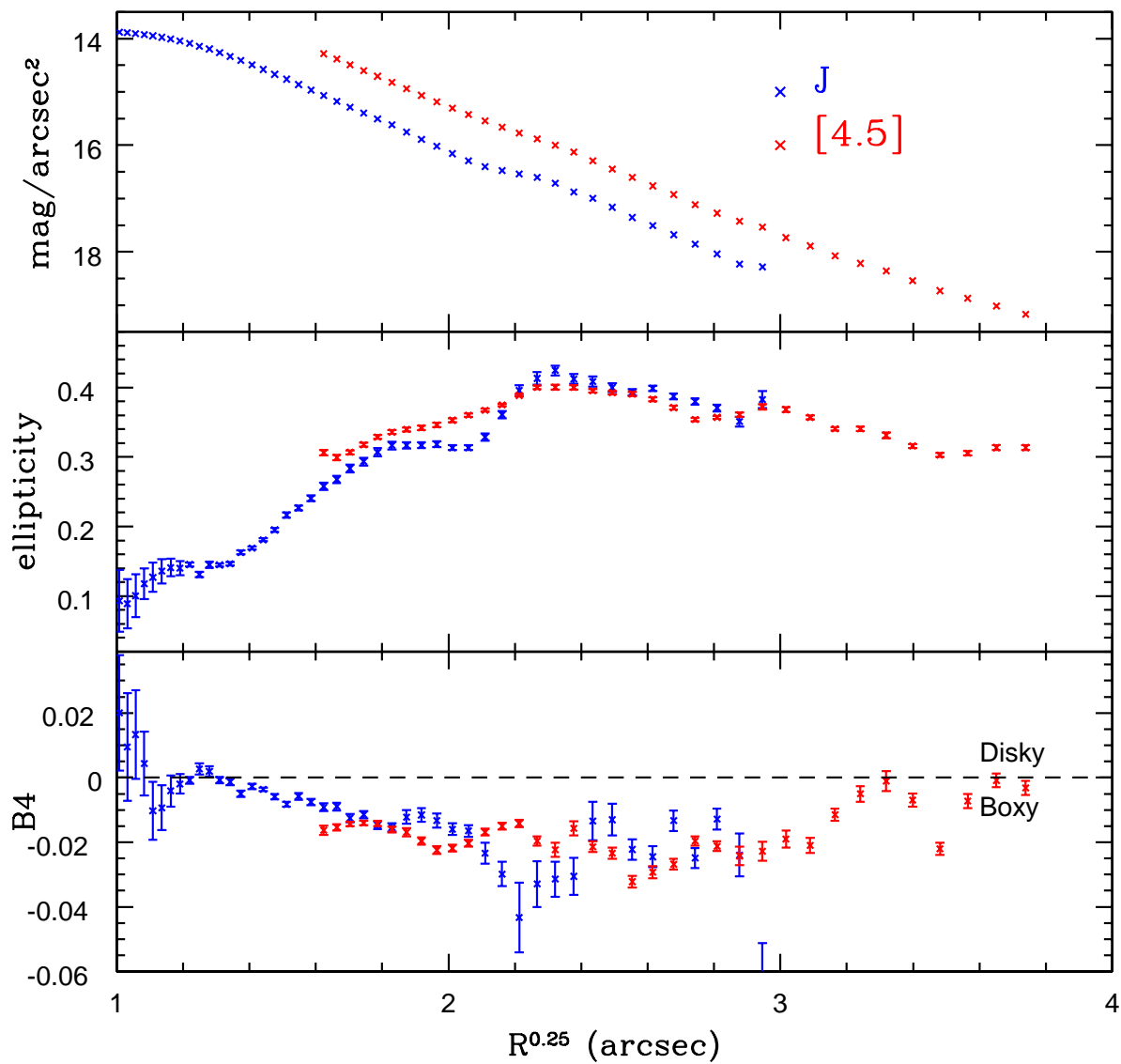}
\caption{Isophotal properties of Maffei 1 in the $J$ (blue 
points) and [4.5] (red points) filters. Radius is measured 
along the major axis, and the error bars show the uncertainties computed 
by {\it ellipse}. B4 -- the coefficient of the fourth order cosine term in the 
Fourier expansion of the isophotes -- has a negative value throughout 
much of the area that is sampled, indicating boxy isophotes.
Saturation restricts isophote fitting in [4.5] to radii $> 6$ arcsec.} 
\end{figure}

	The $J$ and [4.5] surface brightness measurements both follow an R$^{1/4}$ 
profile. The isophotes in $J$ and [4.5] tend to become progressively flatter (i.e. 
trend to higher ellipticities) from $r \sim 6$ arcsec to $r \sim 30$ arcsec. This 
trend reverses at $r \sim 30$ arcsec, with the isophotes becoming progressively 
rounder with increasing radius. The $J$ ellipticities in Figure 1 are broadly 
consistent with those measured by Buta \& McCall (2003) at shorter wavelengths. The 
radial trends in ellipticity in $J$ and [4.5] differ between 10 and 25 arcsec, and in 
Section 3.2 it is shown that this radial interval hosts a population that 
contains very red bright objects.

	B4 distinguishes between boxy (B4 $< 0$) and disky (B4 $> 0$) 
isophote shapes (e.g. Carter 1978). Ellipticals that have boxy isophotes are 
thought to be the result of major mergers (e.g. Khochfar \& Burkert 2005; 
Naab et al. 1999). The frequency of boxy isophotes in galaxies increases towards 
progressively longer wavelengths (e.g. Bureau et al. 2006), and this has been
attributed to the diminished impact of dust absorption towards longer 
wavelengths (e.g. Bureau et al. 2006), which makes obscured structures easier to 
detect. The mix of stellar types that dominate the light at different 
wavelengths may also play a role in the trend between B4 and wavelength. 

	The structural information obtained 
from the $J$ image shows a tendency for B4 to become more disk-like with 
decreasing radius for $r < 10$ arcsec. At these radii the NIR surface brightness 
profile flattens and the ellipticity becomes rounder. 
Buta \& McCall (2003) find that B4 transitions to positive values 
at radii $< 1$ arcsec, as would be expected if the innermost regions of 
Maffei 1 harbored a disk. A central disk could have formed if rotationally 
supported material was channeled into the central regions of the galaxy. 
At radii in excess of a few arcsec the B4 measurements in Maffei 1 tend to be 
negative, but approach zero at radii $> 100$ arcsec. 

\subsection{The Infrared Spectral-Energy Distribution}

	A large fraction of the light from old and intermediate age stellar systems 
in the NIR and MIR originates from the most evolved stars, and so photometry 
in this wavelength region provides information that is 
complementary to photometric measurements at visible wavelengths, where 
sub-giant branch and main sequence stars contribute significantly to 
the integrated light. The $J-K$ color profile constructed from the 2MASS 
surface brightness measurements is shown in the upper left hand corner of Figure 2. 
The radial behaviour of $J-K$ is very similar to that of 
$V-I$ in Figure 24 of Buta \& McCall (1999). $J-K$ is more-or-less constant at 
radii $< 15$ arcsec. There is no evidence for a blue nucleus in the 2MASS data, 
although with a 3 arcsec angular resolution the 
light from a compact nucleus is significantly blurred. 

	The near-constant $J-K$ color at radii $< 15$ arcsec suggests that the 
properties of the brightest red stars do not vary greatly with radius near the center 
of Maffei 1, which is where the B4 coefficient 
trends towards more disky values. Given the near-constant $J-K$ color in 
this interval then absorption features in the NIR spectrum might 
be expected to be radially uniform near the galaxy center. 
It is shown in Section 4 that -- the nucleus aside -- this appears to be the case.

	$J-K$ peaks near $r \sim 20$ arcsec. The red 
color at this radius suggests that there is a larger 
contribution from luminous red stars with lower effective temperatures than 
at other radii. The radial trend in ellipticity obtained from 
the 2MASS $J$ image changes at this radius (Figure 1), suggesting that the changes 
in $J-K$ and the structural properties are linked.

	Maffei 1 becomes progressively rounder (i.e. ellipticity drops) with 
increasing radius at large radii, while the isophotes retain a boxy shape. 
There is also a tendency for $J-K$ to become smaller with increasing radius 
when $r > 25$ arcsec. Such gradients in the outer regions of classical elliptical 
galaxies are typically interpreted as the consequence of radial changes in metallicity 
(e.g. Tamura et al. 2000). 

	While the NIR light is primarily photospheric in origin, 
thermal emission from dust with temperatures of a few hundred K 
in circumstellar shells around AGB stars may contribute significantly to 
the MIR light. The luminosity and SED of this circumstellar emission is 
linked to the initial mass and the metallicity of the host star, 
in the sense that AGB stars with massive, metal-rich progenitors are expected 
to have higher rates of mass loss and hence thicker dust envelopes than 
those that have a lower mass and/or are more metal-poor. Photometry 
in the MIR thus provides a potential probe of the SFH.
The potential utility of emission in the $3 - 5\mu$m regime as a tracer of 
SFH was demonstrated by Davidge (2014) who showed that the model [4.5] luminosity 
function of the brightest AGB stars in M32 obtained by adopting the SFH found 
by Monachesi et al. (2012) from stars near the main sequence turn-off 
is in good agreement with the observations. 

	In order to combine the 2MASS and Spitzer photometric 
measurements, the [4.5] image was convolved with a Gaussian and re-sampled 
to match the angular resolution and sampling of the 2MASS $J$ image. {\it 
ellipse} was then run on the smoothed image, and the resulting $J$--[4.5] color 
curve is shown in the lower left hand panel of Figure 2. The $J$--[4.5] curve has a 
similar shape to the $J-K$ curve at radii $< 25$ arcsec, although at radii $> 25$ 
arcsec there is not a clear color gradient.

\begin{figure}
\figurenum{2}
\epsscale{1.00}
\plotone{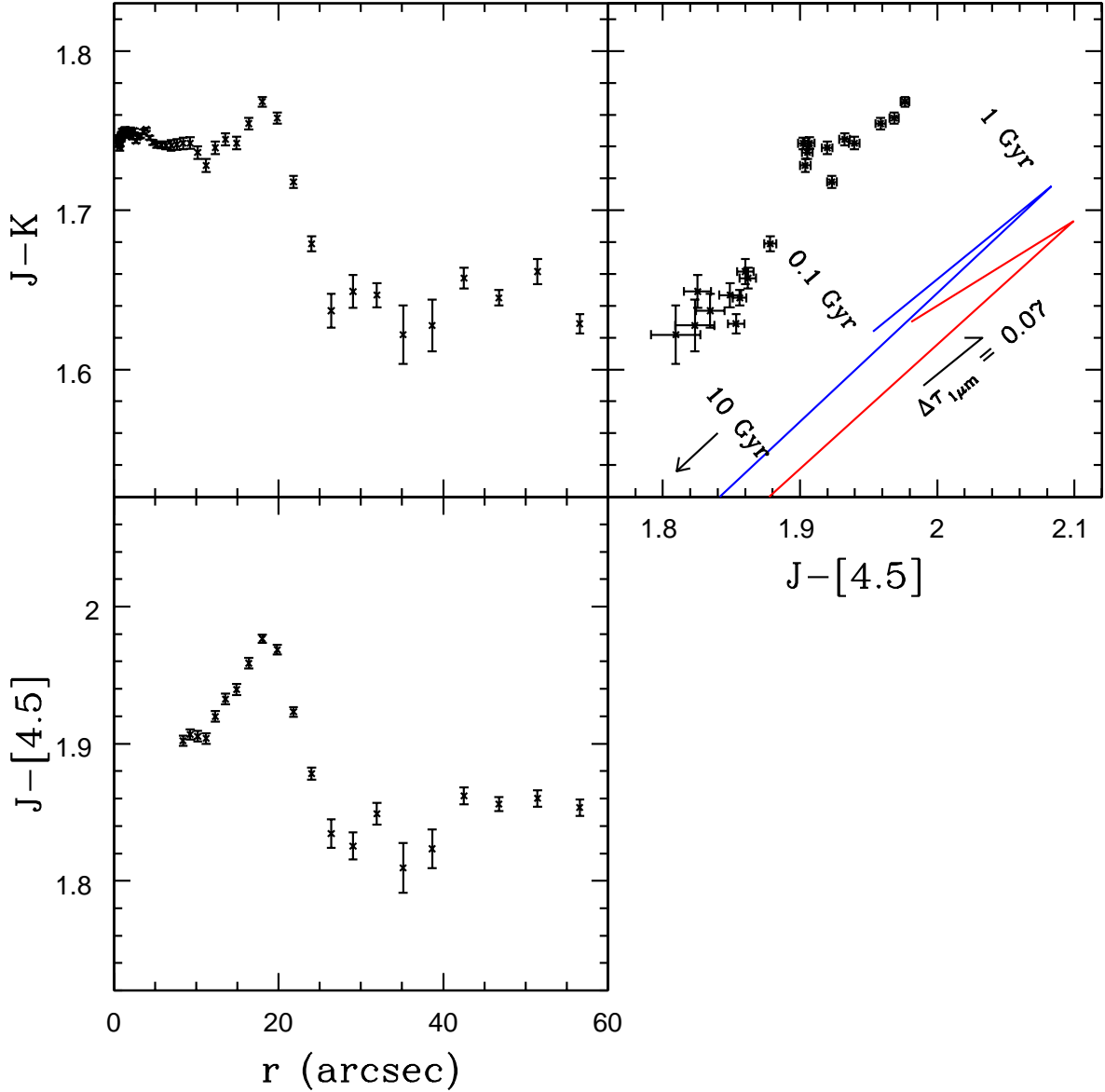}
\caption{Infrared color profiles and two-color diagram. 
Comparisons with Figure 1 indicate that radial changes in 
the color profiles are associated with changes in isophote structure. 
SSP models constructed from the Marigo et al. (2008) isochrones 
are shown in the upper right hand panel. The models assume a Chabrier (2001) 
IMF and metallicities Z=0.020 (blue) and Z=0.040 (red). The 0.1 Gyr and 1 Gyr 
points on the Z=0.020 sequence are indicated. The models have been reddened 
to $\tau_{1\mu m} = 1.69$ using the procedure described in the text. The reddening 
vector in the lower right hand corner of the TCD has a length $\tau_{1\mu m} = 
0.07$, which is the uncertainty in the Fingerhut et al. (2003) 
$\tau_{1\mu m}$ estimate.}
\end{figure}

	Two color diagrams (TCDs) are one means of comparing observed SEDs with 
model predictions. The $(J-K, J-[4.5])$ TCD of Maffei 1 is shown in the upper 
right hand panel of Figure 2. Simple stellar population (SSP) sequences constructed 
from the Marigo et al. (2008) isochrones are compared with the Maffei 1 measurements 
in the upper right hand corner of Figure 2. Stars in an SSP have the same age 
and metallicity; SSPs thus assume an idealized situation in which the 
component stars are the result of a single instantaneous star-forming episode.

	The models were downloaded from the Padova observatory 
website \footnote[5]{http://stev.oapd.inaf.it/cgi-bin/cmd}, and incorporate 
evolution on the thermal-pulsing AGB (TP-AGB) using the procedures described by 
Marigo \& Girardi (2007). Emission from circumstellar dust is included using 
the models described by Groenewegen (2006), adopting a 
$60\%$ silicate$+40\%$ AlOx mix for M stars, and a $85\%$ AMC$+15\%$ SiC 
mix for C stars. While the chemical composition of 
circumstellar dust is uncertain and can affect the SED of the emission, 
this only becomes significant at wavelengths $\geq 5\mu$m 
(e.g. Figure 1 of Groenewegen 2006). 

	A Chabrier (2001) mass function was assumed, and 
the model SSP colors were reddened to $\tau_{1\mu m} = 1.69$ (Fingerhut et 
al. 2003) using the monochromatic extinction curve of Fitzpatrick (1999). 
Extinction coefficients were computed with the York Extinction Solver (McCall 2004), 
which is hosted by the Canadian Astronomical Data Center \footnote[6]
{http://www.cadc-ccda.hia-iha.nrc-cnrc.gc.ca/community/YorkExtinctionSolver/}. 
The ratios of the extinction in each filter to $\tau_{1\mu m}$ assuming 
R$_V = 3.2$ are 0.752 (J), 0.324 (K), and 0.133 ([4.5]). 
The reddening vector, shown in the lower right hand corner of the TCD, parallels 
the isochrones. Models in which TP-AGB evolution is suppressed have much smaller 
$J-[4.5]$ and $J-K$ colors than the sequences shown in Figure 2, and fall outside of 
the color range examined in the figure.

	Incomplete knowledge of the physical processes at work during TP-AGB 
evolution and the emission characteristics of circumstellar dust are among the 
potential sources of uncertainty in the models shown in Figure 2. The rate of mass 
loss prior to the onset of the TP-AGB can also affect the lifetimes of TP-AGB stars, 
and hence the contribution that they make to the total light from an SSP 
(Rosenfeld et al. 2014). This being said, adopting a 
different mineralology for the circumstellar dust does not change significantly 
the location of the models in Figure 2. The placement of the models are also 
only mildly sensitive to R$_V$.

	While there is an offset between the models and observations in the TCD, 
it is encouraging that the model sequences parallel the locus of 
points in Maffei 1, suggesting that the models track {\it differences} in 
the properties of TP-AGB stars well, and that age -- rather than metallicity -- is 
the dominant driver of the radial color variation within 60 arcsec of the nucleus. 
In fact, the locus of points on the TCD defined by observations at $r 
< 20$ arcsec breaks off to the left on the TCD, as predicted by the models 
if the luminosity-weighted age in the inner regions of Maffei 1 
is younger than at $r = 20$ arcsec. Based solely on 
$J-K$ color, the central r $\sim 20$ arcsec region of Maffei 1 has a 
luminosity-weighted age $\sim 1$ Gyr, indicating that a substantial intermediate age 
component is present. Spectroscopic signatures of an intermediate age population 
might then be expected in the central regions of Maffei 1.

\section{SPECTRA}

\subsection{The Light Profile Extracted from the Spectra}

	Before investigating the absorption spectrum of the central regions 
of Maffei 1, we first examine the light profile constructed from 
the IRCS spectrum. Pseudo broad-band photometric measurements were obtained 
by averaging the counts along the dispersion axis between 1.55 and 1.8$\mu$m. 
No attempt was made to reproduce a standard filter response function, although 
the wavelength limits used here roughly approximate the boundaries of the $H$ filter. 
Given the slit orientation, the resulting light profile samples the minor axis.

	The light profile constructed from the spectrum is shown in Figure 3. There 
is a pronounced central peak that has a characteristic width of 0.1 -- 0.2 arcsec. 
The solid line in the right hand panel is the Buta \& McCall (2003) 
profile of Maffei 1 obtained from WFPC2 observations in F814W, and 
translated onto the minor axis by applying the ellipticities in 
their Figure 9. The Buta \& McCall (2003) profile has 
also been shifted along the y axis to match the Subaru profile at log(r) $= 0.0$. 
There are only small differences between the Subaru and HST light profiles, 
suggesting that the RAVEN-corrected PSF has a characteristic width that is 
not vastly broader than that produced by HST. 

\begin{figure}
\figurenum{3}
\epsscale{1.00}
\plotone{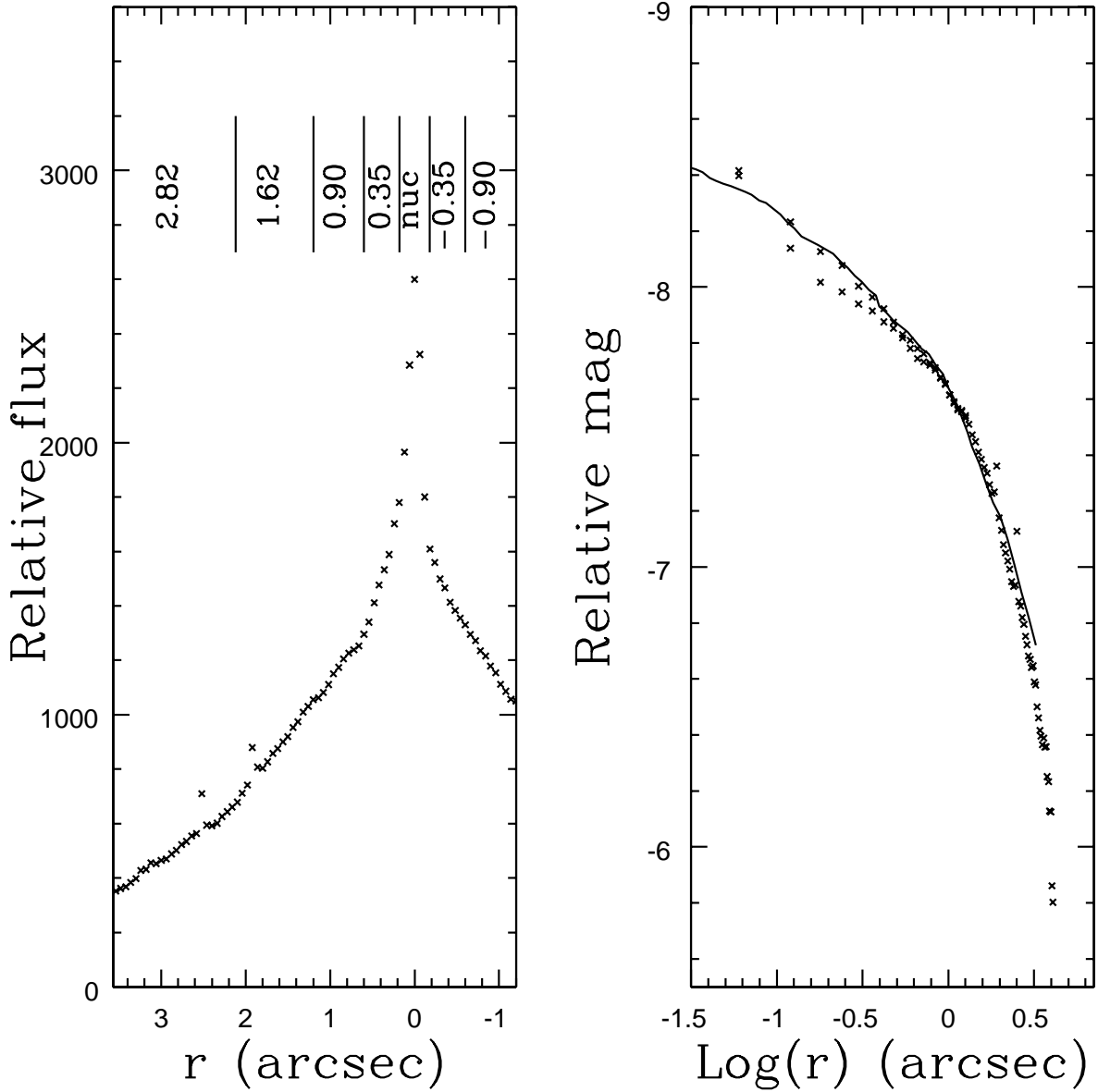}
\caption{NIR light profile of Maffei 1, which was obtained by averaging the IRCS 
spectra along the dispersion axis in the wavelength interval $1.55 - 1.8\mu$m. $r$ is 
the distance from the nucleus along the minor axis. The radial binning intervals 
for spectral extraction are indicated, and the mean angular offets from the nucleus 
are specified. The right hand panel shows the light profile in log-log co-ordinates. 
The solid line is the F814W profile from Buta \& McCall (2003), corrected to fall 
along the minor axis by applying the ellipticities in their Figure 9. The profile 
was then shifted vertically to match the IRCS profile at log(r) $= 0.0$.}
\end{figure}

\subsection{The Radial Strengths of Absorption Features}

	Spectra were extracted in various radial intervals to examine the 
uniformity of the stellar content. The binning intervals, 
indicated in the left hand panel of Figure 3, 
were selected based on angular resolution and S/N ratio. The extracted 
spectra were convolved with a gaussian along the dispersion axis to simulate the 
higher velocity dispersion in the nuclear spectrum, and all spectra were 
shifted into the rest frame to correct for rotation within Maffei 1.

	The mean of the extracted circumnuclear spectra in the --0.90, 0.90, 
and 1.62 arcsec intervals in Figure 3, referred to here as the circumnuclear 
spectrum, and of the nucleus are shown in Figure 4. Spectra of K and 
M giants from the compilation of Rayner et al. (2009), processed to match 
the effective spectral resolution and wavelength sampling of 
the Maffei 1 observations, are also shown in this figure. 
The stars in the Rayner et al. (2009) compilation are in the 
Galactic disk, and so likely have solar metallicities and chemical mixtures. 
All spectra in Figure 4 have been divided by a continuum function.

\begin{figure}
\figurenum{4}
\epsscale{0.90}
\plotone{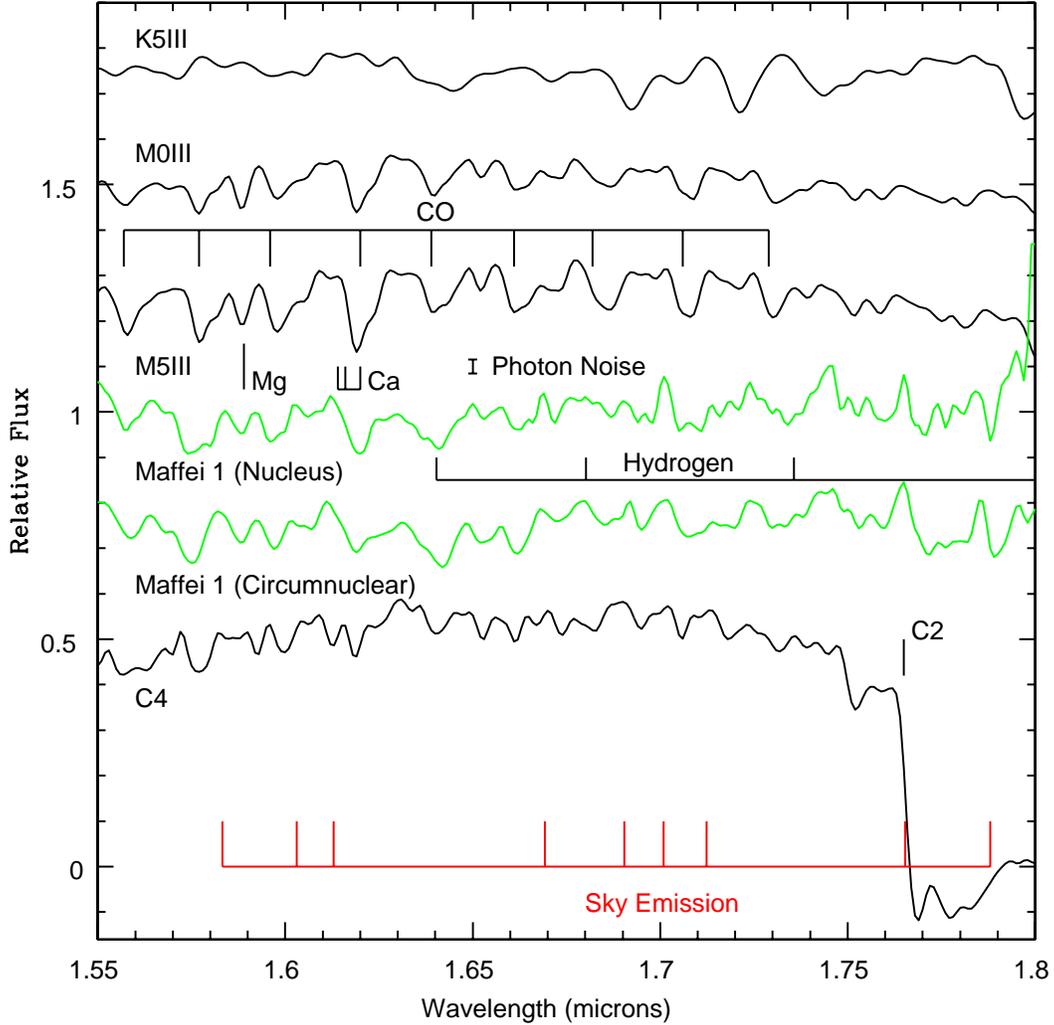}
\caption{Spectra of the nucleus and circumnuclear regions of Maffei 1 are 
compared with spectra of bright Galactic giants from Rayner et al. (2009). The 
`Circumnuclear' spectrum is the average of the extracted spectra in the 
--0.90, 0.90, and 1.62 intervals indicated in Figure 3. The error bar 
next to the `Photon Noise' label shows the $\pm 2\sigma$ uncertainty in 
the nucleus spectrum as estimated from the mean number of photons 
from the galaxy and sky. The stellar spectra have been re-sampled and 
smoothed to simulate the effective spectral resolution of the Maffei 1 
spectra, including the central velocity dispersion. The Maffei 1 spectra have been 
shifted to the rest frame. Various molecular and atomic features are marked, as are 
the locations of strong telluric emission lines. 
Brackett H emission is not detected in the Maffei 1 spectra. There is a break in the 
Maffei 1 spectra near $1.76\mu$m, which is the wavelength of the C$_2$ bandhead. 
While there is a feature near the C$_2$ bandhead with the 
same wavelength as a night sky line, it has the same strength in the 
nucleus and circumnuclear spectra, which would not be expected for a residual 
sky feature. This bump also appears in model spectra presented in Figure 7, 
indicating that it is not an artifact of poor sky subtraction, but is inherent to 
the spectrum of Maffei 1.} 
\end{figure}

	Various molecular and atomic absorption features are marked in Figure 4. 
The second overtone bands of CO are the dominant features in the H-band 
spectra of late-type giants. However, some atomic transitions are also detected at 
this resolution, and the deepest absorption feature in the H-band spectra of M giants 
and the nucleus of Maffei 1 is a blend of Ca and CO. 
While Buta \& McCall (2003) found centrally-concentrated H$\alpha$ emission, 
Brackett hydrogen emission is not seen in the Maffei 1 spectra. 

	There is reasonable agreement between the spectra of Maffei 1 and 
the M0III template between 1.55 and 1.65$\mu$m. However, at longer wavelengths the 
quality of agreement degrades. A similar situation is seen in the $H-$band 
spectrum of the lenticular galaxy NGC 5102, which Miner et al. (2011) compare with the 
spectrum of M32 in their Figure 1. The Ballick-Ramsey C$_2$ band head near $1.76\mu$m 
is a prominent C star signature, and Miner et al. (2011) detect it in their spectrum 
of NGC 5102. The Maffei 1 spectra show a break near 1.76$\mu$m that is similar to 
that in the Miner et al. (2011) NGC 5102 spectrum, hinting that the stellar content 
of the central regions of Maffei 1 may be more closely related to that of the 
center of NGC 5102, which contains a large population with an age $\sim 1$ Gyr 
(Davidge 2015), than to that of older (e.g. 3 -- 4 Gyr) systems like M32.
If -- as suggested by the NIR and MIR colors discussed in Section 3 -- the NIR light 
is dominated by a large population with an age $\sim 1$ Gyr then C stars might 
be present in Maffei 1.

	The uniformity of line strengths along the 
IRCS slit is examined in Figure 5. The nuclear spectrum of Maffei 1 from Figure 
4 is re-plotted in the top panel of Figure 5, while the 
differences between the nuclear spectrum and the extracted spectra 
in the circumnuclear environment are shown in the lower panel. 
There are similarities among the differenced spectra in Figure 5, 
suggesting that the brightest red stars in the circumnuclear region are uniformly 
mixed with radius. Many of the strongest residuals coincide with the CO band heads, 
and the residuals are such that these features are stronger in the 
nucleus than in the circumnuclear region. The nucleus thus contains a 
stellar mix that differs from that in its immediate surroundings.

\begin{figure}
\figurenum{5}
\epsscale{0.90}
\plotone{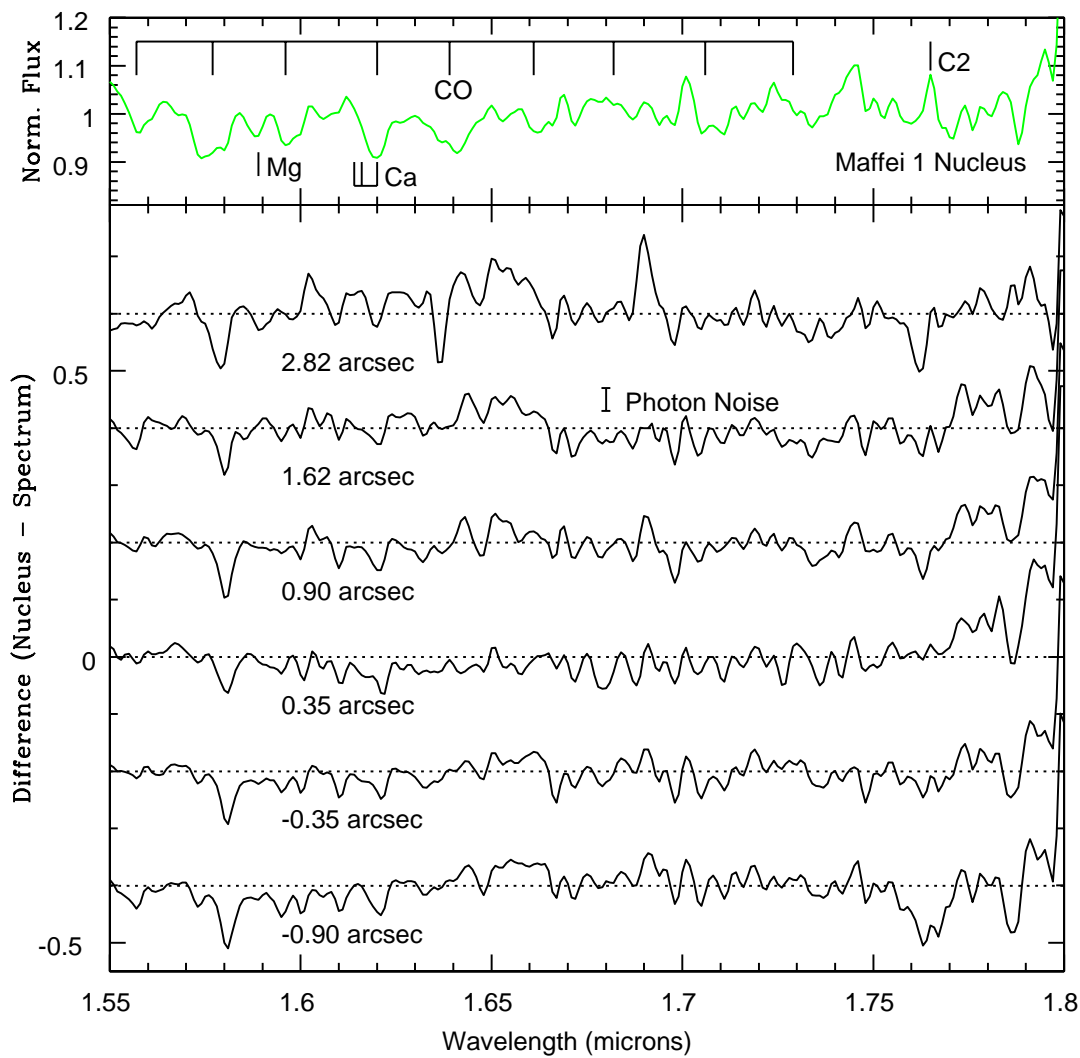}
\caption{Differences between the nuclear and circumnuclear spectra. The 
nuclear spectrum of Maffei 1 from Figure 4 is shown in the top panel. 
The differences between the spectra have been shifted along the y-axis for the 
puposes of display, and the dotted lines indicate zero levels in the differences. 
The error bar next to the `Photon Noise' label shows the typical $\pm 2\sigma$ 
scatter calculated from the mean number of photons and the contribution made 
by the sky in the various radial intervals. This noise does 
not vary greatly from interval to interval, and so only 
one error bar is shown. The residuals do not show systematic trends 
with radius, suggesting that the bright red stellar content is well mixed 
throughout the circumnuclear region. Large residuals occur in the vicinity of 
many of the CO bandheads, in the sense that these features tend to be stronger in the 
nucleus than in the circumnuclear region. This suggests that the mix of bright red 
stars in the nucleus differs from that in the circumnuclear region.} 
\end{figure}

\subsection{Comparison with Models}

	The stellar content of the central regions of Maffei 1 
is explored in this section by making comparisons with model spectra. The 
spectra of the central regions of Maffei 1 are found to contain signatures 
of an intermediate age population that contributes 
significantly to the NIR light, as might be expected given the broad-band IR 
colors discussed in Section 3. Comparisons are made with two sets of published 
models (Section 4.3.1) and with models constructed from 
the Rayner et al. (2009) spectral library (Section 4.3.2).
The models are restricted to SSPs, and a detailed 
breakdown of the SFR as a function of time in Maffei 1 will best be done with 
spectra that span a broader wavelength interval than considered here.

\subsubsection{Comparisons With Published Models}

	There is only a modest body of published NIR model spectra that can be 
compared with the observations, and these models tend to have 
a low spectral resolution. For the present work, model spectra from the Bag of 
Stellar Tricks and Isochrones (BaSTI; Cordier et al. 2007) and Maraston (1998; 2008) 
are considered. These models have a spectral resolution $\sim 170$. 
While having a low resolution, these models have the merit of covering a broad 
range of ages and metallicities, and allow discernible age signatures to be examined. 
They thus provide a guide for identifying promising features for more detailed 
examination. The Cordier et al. (2007) and Maraston (1998; 2005) models reproduce 
some of the broad-band photometric properties of star clusters (e.g. Salaris et 
al. 2014), which is an important test of their validity.

	Solar metallicity SSP models are shown in 
the left hand (BaSTI) and right hand (Maraston) panels of Figure 6. 
A Chabrier (2001) IMF was used to construct the BaSTI models, 
while a Kroupa (2001) IMF was adopted for the Maraston models. This difference 
in IMFs should not be a concern when considering SSPs at these wavelengths 
as a large fraction of the light near $1.6\mu$m originates from 
stars that are highly evolved, and thus span a small mass range. 

\begin{figure}
\figurenum{6}
\epsscale{0.90}
\plotone{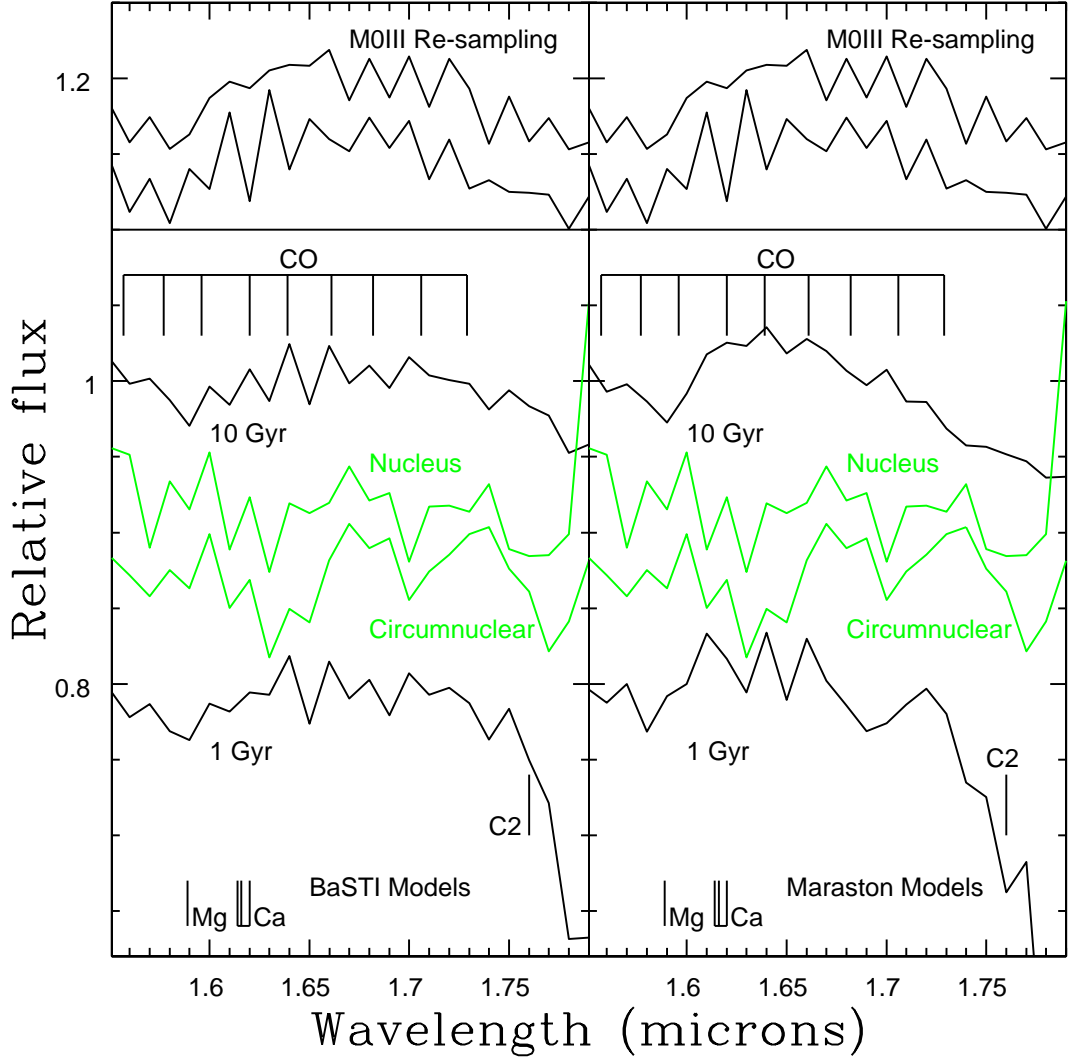}
\caption{Comparing solar metallicity models from the BaSTI (lower left hand panel) 
and Maraston (lower right hand panel) libraries. The impact of binning on 
the model spectra is examined in the top panels, where an M0III 
spectrum from Rayner et al. (2009) is shown after applying starting wavelengths 
for binning that are separated by half a resolution element. The largest 
differences occur between 1.6 and 1.65$\mu$m. The Maffei 1 nuclear and circumnuclear 
spectra, binned to match the spectral resolution of the models, are shown in 
green. The absorption features that are seen at this resolution are predominantly 
molecular in origin. There are obvious differences between the 1 and 10 Gyr models. 
In particular, the $1.75 - 1.80\mu$m interval in the 1 Gyr model is dominated 
by C$_2$ absorption, and this feature is not seen in the 10 Gyr models. 
Evidence of C$_2$ absorption is also seen in the Maffei 1 spectra.}
\end{figure}

	The low spectral resolution of the models causes deep absorption 
features to be blended together, and so the starting wavelength for binning 
can influence the appearance of the models. The affect of the starting 
wavelength for binning at this resolution is investigated in the top panels 
of Figure 6, where spectra of an M0III star from Rayner et al. (2009) are 
compared. The spectra in the top panel show the result of 
changing the initial wavelength for binning by 
one-half of a resolution element, thereby showing the largest difference 
that might be expected. There are significant differences 
between the two spectra. While the locations of absorption features in the 
two spectra are similar, the depths of absorption features depend on 
the starting wavelength. The largest differences are evident in 
the $1.6 - 1.65\mu$m wavelength region.

	There are noticeable differences between the BaSTI and Maraston models. 
In addition to binning-related effects, these are likely due 
to differences in the source spectra from which the models 
are constructed and the methods used to construct the evolutionary sequences 
upon which the models are based. The techniques used to model the TP-AGB in the BaSTI 
models are described by Cordier et al. (2007). Evolution prior to the onset of thermal 
pulses is tracked with a conventional stellar structure code. The TP-AGB is modelled 
using the techniques described by Iben \& Truran (1978), and includes 
hot-bottom burning and mass loss. The effective temperature at each point on 
the TP-AGB is computed from relations that link total stellar mass and 
the mass of the carbon-oxygen core.

	In contrast to the BaSTI models, the Maraston (1998; 2005) models use the fuel 
consumption theorum to determine the contributions made by stars at different 
evolutionary stages. Empirically-based calibrations then link evolutionary states 
to observational properties. There is evidence that 
models based on the fuel consumption theorum may have 
difficulty reproducing the numbers of TP-AGB stars in intermediate age LMC clusters 
(Girardi et al. 2013). The fractional contribution made by C stars to the 
integrated light is determined using the metallicity-related arguments 
outlined by Renzini \& Voli (1981). C stars are formed at all metallicities, 
but with a higher frequency and over a broader range of masses towards lower 
metallicities (Figure 12 of Maraston 2005).

	Percival et al. (2009) describe the source spectra used to 
construct the BaSTI models. Model atmospheres are used for stars warmer than 
3500 K, while empirical spectral libraries are adopted for stars with 
T$_{eff} < 3500$K, including C stars. Real stellar spectra are used for 
the cool stellar component to avoid uncertainties in model atmospheres 
at low effective temperatures. Maraston (2005) also 
combines model atmospheres and empirical stellar spectra, but uses atmosphere 
models constructed from a different code than that employed by Percival et al. (2009).

	The second overtone CO bands produce 
a sawtooth pattern that is most noticeable in the 10 Gyr BaSTI model. The 1 Gyr 
and 10 Gyr models in both panels of Figure 6 have different characteristics, 
and the CO bands in the Maraston models are noticeably deeper in the 1 Gyr model 
than in the 10 Gyr model. This age dependence is due to the larger contribution 
made by luminous, cool AGB stars to the integrated NIR light in the younger model.

	While there are substantial differences between the BaSTI and Maraston models, 
there are also common features. A noteable difference between 
the 1 and 10 Gyr models in both sets of models is seen at wavelengths longward of 
$1.75\mu$m, and is due to the appearance of the 
Ballick-Ramsey C$_2$ band in the 1 Gyr models. A corresponding feature is 
also seen in the nuclear and circumnuclear spectra of Maffei 1, which 
are also shown in Figure 6. The Maffei 1 spectra were processed to match the 
resolution of the models, using a starting wavelength for binning that produced the 
best visual agreement with the models. The appearance of the C$_2$ feature in 
the Maffei 1 spectra is robust in terms of the initial wavelength for binning.

	C stars are expected in populations that span only a limited range of 
metallicities and ages, and models of single star evolution predict that C stars 
do not form in populations with an age of 10 Gyr. In the disk of M31 the C/M star 
ratio decreases towards smaller galactocentric radius (e.g. Brewer et al. 1996; 
Boyer et al. 2013). While this trend is usually interpreted as a metallicity 
effect, it could also reflect a dearth of star-forming activity during an epoch 
that would result in a population of C stars at the present day. Battinelli \& Demers 
(2005) find a clear trend between C star frequency and metallicity in nearby galaxies.

	As a moderately large elliptical galaxy, Maffei 1 might be expected 
to have a solar or higher mean metallicity. This does not preclude the formation of 
C stars, as models of C star evolution predict that they will form at solar 
metallicities, albeit with a mass range that decreases with increasing metallicity 
(e.g. Karakas 2014; Mouhcine \& Lancon 2003). The chances of finding other large 
ellipticals like Maffei 1 that have C star signatures are thus lower than if a sample of 
(lower metallicity) dwarf ellipticals were compared, as there is a narrower range of SFHs 
that will produce significant numbers of C stars in the more massive galaxies.

\subsubsection{Comparisons With New Models}

	Neither the BaSTI nor Maraston models give a satisfactory match 
to the Maffei 1 spectra, even outside of the $1.6 - 1.65\mu$m interval where 
binning effects are most pronounced. Thus, given the level of 
agreement between the Maffei 1 observations and the Rayner et al. (2009) 
standard stars in Figure 4, a modest suite of model spectra were constructed using 
the Rayner et al. (2009) stellar library. The models are based on the evolutionary 
sequences from Marigo et al. (2008), which include advanced stages of 
evolution that follow prescriptions discussed by Bertelli et al. (1994), 
Girardi et al. (2000), and Marigo \& Girardi (2007). The model spectra are restricted 
to solar metallicity given that the stars in the Rayner et al. (2009) 
sample are in the solar neighborhood. Spectral type $vs.$ T$_{eff}$ 
relations from Pecaut \& Mamajek (2013) and Richichi et al. (1999) were used to 
assign spectral types to the various evolutionary stages. A Salpeter (1955) IMF was 
assumed. The contribution from faint MS stars is modest when compared with highly 
evolved objects, and so the use of a different mass function will not change 
greatly the results. The baseline models assume oxygen-dominated photospheres, 
although it is shown below that the inclusion of C star spectra improve the agreement 
with the observations.

	The models were smoothed and re-sampled to match the spectral resolution of 
the IRCS spectra, and the results are compared with the Maffei 1 spectra in Figure 7. 
The absorption features in the 0.5 and 1 Gyr models are deeper than in the 10 Gyr 
models, reflecting the larger contribution made by very luminous late-type 
AGB stars to the integrated NIR light during intermediate epochs. 
While the depths of absorption features in the Maffei 1 spectra are better matched 
by the 0.5 and 1 Gyr models than the 10 Gyr model, it should be kept in mind that 
there is an age-metallicity degeneracy. As a result, 10 Gyr models 
with super-solar metallicities would likely yield better agreement with the Maffei 1 
spectra than the solar metallicity 10 Gyr models.

\begin{figure}
\figurenum{7}
\epsscale{1.00}
\plotone{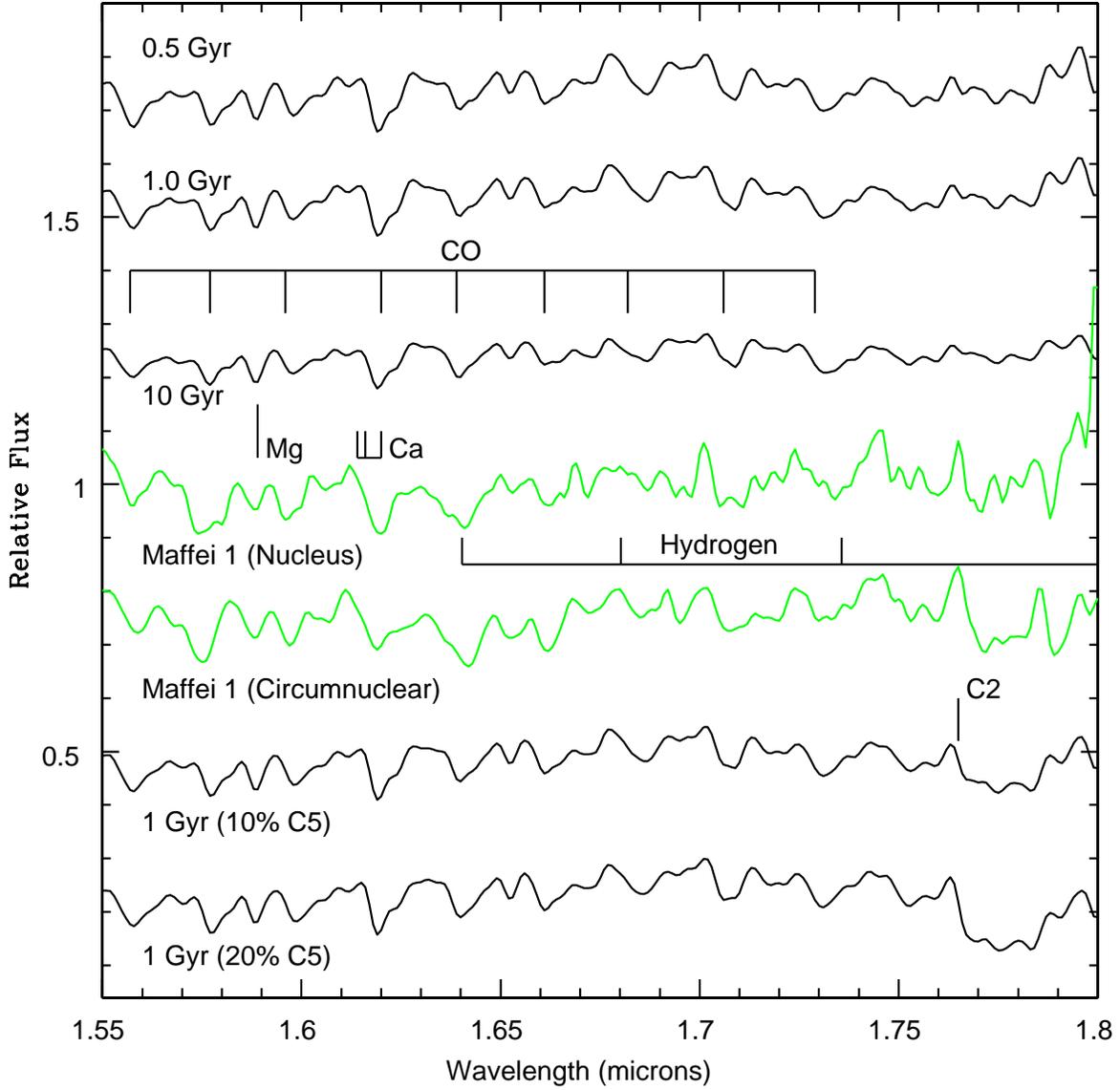}
\caption{SSP model spectra constructed from the Rayner et al. (2009) stellar library. 
The Maffei 1 spectra are shown in green. The models are based on the Marigo et 
al. (2008) solar metallicity evolutionary sequences, with advanced stages of evolution 
modelled according to Bertelli et al. (1994), Girardi et al. (2000), and Marigo \& 
Girardi (2007). The models have been re-sampled and smoothed to match the spectral 
resolution of the Maffei 1 observations, including the central velocity dispersion. 
A Salpeter (1955) IMF is assumed. The bottom two spectra show the result of adding a 
C5 spectrum to the 1 Gyr model, with this spectral type accounting for 10\% and 20\% 
of the total light. The Maffei 1 circumnuclear spectrum in the $1.76 - 1.80\mu$m 
interval is better matched by the models that include C stars.}
\end{figure}

	The baseline models in Figure 7 do not match the Maffei 1 spectra in the 1.75 
-- 1.80$\mu$m interval, which is the wavelength region containing the Ballick-Ramsey 
C$_2$ bands. Models were thus investigated in which the spectrum of a C5 star was 
added. This spectral type is the most common among C stars in the Galactic disk 
(Alksnis et al. 2001). Models in which a C5 star contributes 10\% and 20\% of the 
total light in $H$ are shown in Figure 7. 

	The differences between the circumnuclear spectrum and the 
various models are shown in Figure 8. The addition of the C5 spectrum 
yields a better match to the circumnuclear spectrum 
than the baseline models. The dispersion of the residuals over all wavelengths in the 
circumnuclear spectrum is at a minimum if 20\% of the light comes from C stars. 
For comparison, the minimum dispersion for the nuclear spectrum 
occurs when there is a 10\% contribution from C stars. The residuals are 
much larger when C stars are excluded from the models.

\begin{figure}
\figurenum{8}
\epsscale{1.00}
\plotone{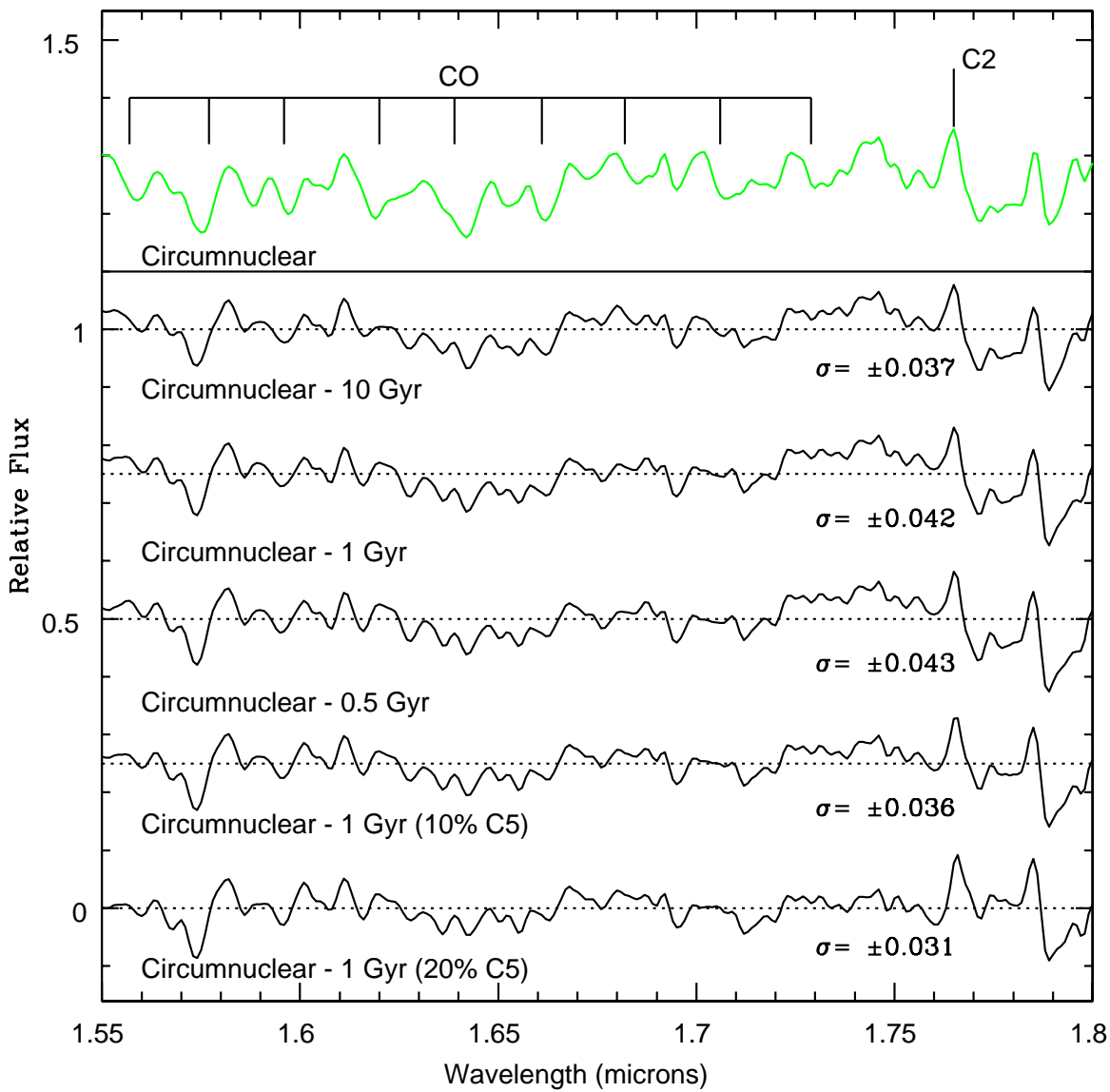}
\caption{Differences between the Maffei 1 circumnuclear spectrum and the models shown 
in Figure 7. The circumnuclear spectrum from Figure 4 is shown in the top panel. 
The differential spectra have been shifted along the y-axis for display purposes, and 
the dotted lines indicate zero residual levels. The dispersions 
in the residuals between 1.7 and $1.8\mu$m, where the difference between 
the models with and without C stars is greatest, are indicated. The dispersions 
measured in the 1.55 to $1.8\mu$m interval follow the same trend as those measured 
between 1.7 and $1.8\mu$m. The smallest dispersion in the residuals occurs when 
the 1 Gyr $+$ 20\% C5 model is subtracted from the circumnuclear spectrum.}
\end{figure}

\section{DISCUSSION \& SUMMARY}

	The stellar content in the central few arcsec of the nearby elliptical galaxy 
Maffei 1 has been investigated using NIR spectra and archival IR images. The spectra 
were recorded with RAVEN$+$IRCS on the Subaru Telescope. 
The nucleus of Maffei 1 has an angular size of $\sim 0.15$ arcsec 
FWHM in these data, corresponding to $\sim 2.4$ parsecs at the distance of Maffei 1. 
The good angular resolution delivered by RAVEN is demonstrated 
in Figure 3, where it is shown that the light profile constructed from the IRCS 
spectra matches that obtained from the HST. These data thus demonstrate the 
ability of RAVEN to deliver scientifically interesting angular resolutions. 

	As one of the nearest large elliptical galaxies, Maffei 1 is an important 
laboratory for understanding the evolution of early-type galaxies in group 
environments. The central regions of Maffei 1 harbor important clues of its past 
evolution. The IR colors of the central 20 arcsec are redder than at larger radii, 
suggesting that luminous AGB stars contribute a higher fraction of the light 
within 20 arcsec of the nucleus than at larger radii. In addition, 
the trend between ellipticity and radius changes markedly 
along the major-axis near $\sim 20$ arcsec. The shape of the isophotes as 
gauged by B4 -- the coefficient of the fourth order cosine term in the 
Fourier expansion of the isophotes -- stays more-or-less constant at $ r > 20$ arcsec. 
However, at radii $< 10$ arcsec, B4 approaches and exceeds 0, thereby hinting that 
a disk-like structure is present near the galaxy center. It thus appears 
that the central 20 arcsec of Maffei 1 differs from the galaxy at larger 
radii in terms of stellar content and isophotal properties. 

	Fingerhut et al. (2003) find that the equivalent 
width of H$\beta$ absorption in Maffei 1 is 3.6\AA\ , which 
is indicative of an intermediate age population (e.g. Worthey 1994).
In the present study additional evidence for an intermediate age population 
is found in the form of (1) red IR colors near the galaxy center, and (2) an 
absorption feature longward of $1.76\mu$m that is identified with C$_2$. Comparisons 
with models indicate that the depth of the C$_2$ feature can be reproduced 
if 10 -- 20\% of the $H-$band light originates from stars of spectral type C5, 
which is the most common spectral type for C stars in the solar neighborhood. 
The $J-K$ color of the central regions of Maffei 1 suggests an 
age of $\sim 1$ Gyr. While it is not clear what uncertainty should be assigned 
to age estimates made from integrated NIR light and spectra, 
C stars are expected to make their largest contribution to the 
light from solar metallicity SSPs at ages $\sim 1$ Gyr 
(Figure 12 of Maraston 2005), which is consistent with the age estimated from 
integrated light. Photometry and spectroscopy thus yield a qualitatively 
consistent picture of the stellar content of the central regions of Maffei 1. 

	The origins of the gas that sparked a central star-forming episode in 
Maffei 1 $\sim 1$ Gyr in the past is a matter of speculation. One possible 
source of gas is an interaction with a gas-rich companion. If the companion had a 
lower mass than Maffei 1 then the accreted gas would have formed stars with 
a lower metallicity than older stars in Maffei 1, which presumably formed as part 
of a more massive system. The metallicity distribution function of intermediate age 
stars near the center of Maffei 1 would then differ from that of older stars in 
the galaxy. If lower metallicity gas did fuel star formation near the 
center of Maffei 1 then it would broaden the mass range over which C stars could form.

	A merger a few Gyr in the past may have left 
other observational signatures, although detecting these may prove difficult given 
the location of Maffei 1 along the Galactic disk plane. 
There is no evidence for shells at large radii around Maffei 1, although these 
would be difficult to detect given the substantial line of sight extinction. 
Residual dust lanes throughout the main body of Maffei 1 might also be expected 
after a merger with a gas-rich system, although if a merger occured more 
than a few crossing times in the past then any residual dust may have been 
consumed and/or dispersed by star formation. Buta \& McCall (2003) conclude that 
the complex network of dust absorption that is visible in their deep HST images 
is probably Galactic in origin. Finally, the blue globular clusters found by 
Davidge \& van den Bergh (2005) may be remnants of the elevated levels of star 
formation that would have resulted from a gas-rich merger.

	Maffei 1 is in close physical proximity to the barred spiral galaxy Maffei 2 
(e.g. McCall 2014). The LGA image of Maffei 2 shows a plume of stars extending to the 
south of Maffei 2, which is suggestive of a tidal tail. If there was an interaction 
between Maffei 1 and Maffei 2 then the former galaxy -- which is intrinsically 
brighter and hence should be more massive -- may have 
received gas from the latter. Such an encounter between 
Maffei 1 and Maffei 2 might also have triggered large scale star 
formation in Maffei 2. Wu et al. (2014) present NIR CMDs of the outer 
regions of Maffei 2 and the late type spiral galaxy IC 342, which may be 
part of the same group as Maffei 2. There is evidence that IC 342 hosts star burst 
activity (e.g. Isizuki et al. 1990). When compared with IC 342, Maffei 2 has only 
a modest number of AGB stars, and these have a lower peak brightness. This 
suggests that there were not elevated levels of star-forming activity in 
the outer regions of Maffei 2 during intermediate epochs.

	Some observational predictions can be made based on 
the conclusions reached in this study. The $J-K$ color in 
the central 20 arcsec of Maffei 1 is more-or-less constant, 
indicating that bright red stars are well-mixed throughout this part of the galaxy. 
Long slit spectra of Maffei 1 at visible wavelengths should then show signatures of 
deep H$\beta$ absorption out to a major axis radius of 20 arcsec, at which point 
evidence may be found for an even younger population in the circumnuclear ring. 
Spectroscopic signatures of C stars should also be found in NIR spectra throughout the 
central 20 arcsec. Finally, C stars will contribute an even larger fraction of the 
light at longer wavelengths than those sampled here by IRCS. One of the most 
prominent features in the absorption spectrum of C stars occurs in the 
$2.9 - 3.2\mu$m wavelength interval, and is attributed to HCN$+$C$_2$H$_2$. 
Spectra of the central regions of Maffei 1 at this wavelength should show this 
feature. 

\acknowledgements{Thanks are extended to the referee -- Marshall McCall -- for 
comments that helped improve the paper. This work has made use of 
BaSTI web tools.}

\end{document}